# Feasibility Study of Social Media for Public Health Behaviour Changes

Oluwaseun Ajao, Anna Jurek, Aisling Gough, Ruth Hunter, Eimear Barrett, Gary McKeown,
Jun Hong, Frank Kee
Queen's University Belfast
BT7 1NN Belfast UK

**Abstract**

Social networking sites such as Twitter and Facebook have been shown to function as effective social sensors that can "feel the pulse" of a community. The aim of the current study is to test the feasibility of designing, implementing and evaluating a bespoke social media-enabled intervention that can be effective for sharing and changing knowledge, attitudes and behaviours in meaningful ways to promote public health, specifically with regards to prevention of skin cancer. We present the design and implementation details of the campaign followed by summary findings and analysis.

**Background**

Research has shown that social networks can mediate the transmission of healthy and unhealthy behaviors in populations; either through selection (Centola, 2010, 2011) or influence (Cha et al, 2010) Social Media (SM) platforms have also been shown to transmit moods, feeling and behaviours (Naveed et al, 2011). There are several studies that have shown the effectiveness of social media in terms of behavioural changes in public health interventions such as in physical activity (Cavallo et al, 2012), sexual health (Bull et al, 2012) and risky sexual behaviours (Jones, Baldwin, & Lewis, 2012). To the best of our knowledge our study is one of the first to use Twitter and Facebook social networking platforms to study public health behaviour while raising awareness about skin cancer and its prevention.

**Objective**

The study aimed to address the following research questions to support the feasibility assessment: (1) Does SM constitute an acceptable means for delivering public health information in the target population? (2) Are people willing to share personal issues (e.g. health behaviours or attitudes) across a SM platform? (3) What type of SM communication would attract the attention of the target population? (4) Are individuals, organizations, celebrities more likely to tweet or re-tweet messages related to the public health campaign? (5) What are the key factors that motivate users to share messages amongst themselves?

**Table 1. Household survey statistics of social media usage in Northern Ireland**

| SM Site | Used Frequently/ Sometimes | % | Infrequently/ Not At All | % |
|---|---|---|---|---|
| Facebook | 402 | 53% | 350 | 47% |
| Twitter | 118 | 16% | 634 | 84% |
| SnapChat | 112 | 15% | 640 | 85% |
| Instagram | 99 | 13% | 653 | 87% |
| Google+ | 80 | 11% | 672 | 89% |
| Pinterest | 52 | 7% | 700 | 93% |
| LinkedIn | 36 | 5% | 716 | 95% |
| Other | 27 | 4% | 725 | 96% |
| Vine | 14 | 2% | 738 | 98% |
| Tumbler | 13 | 2% | 739 | 98% |

**Methods**

We began by conducting a survey of 752 households to understand SM usage amongst people in Northern Ireland–the study's target population. We found Facebook and Twitter to be the two most popular platforms as shown in Table 1. To prepare for the two main phases of the intervention we chose hashtags which broadly differentiated skin cancer awareness from skin surveillance messages respectively. The first Phase which ran from the 1st May – 15th July 2015 contained messages with the #SkinSmartNI, #SkinSavvyNI hashtags. The second Phase ran from 1st August - 30th September 2015 and used the hashtag #KnowYourSkinNI. We chose influencers (including radio, TV weather presenters and celebrities such as music artistes) who we hoped would help diffuse our messages. A coordinated SM event promoting the campaign – a Thunderclap – was designed and then delivered on 1st September 2015 with the aim of creating a trending online meme of the various hashtags used. Figure 1 shows the five message types posted - *shocking, story, informative, opportunistic and humorous.*

To effectively capture the Twitter data we chose to subscribe to a data provider for the provision of 100% access to the Twitter firehose while Facebook data collected from the analytics dashboard was sufficient for this purpose. However, due to privacy concerns, analysis of Facebook data is limited and beyond the scope of this current paper. JSON data was parsed into CSV and an SQL database for analysis.

Table 2. Popular tweets by impressions, engagements and retweets

| | | Impressions | Engagements | Retweets |
|---|---|---|---|---|
| **Most "impressions" and most engaging tweet in Phase1 (shocking/disgust)** | | | | |
| 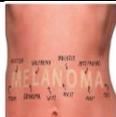 | @xyz One for the ladies, pls help spread the word- Melanoma is 30% more common for women #SkinSavvyNI #eek | 1,349 | 811 | 11 |
| **Most "impressions" tweet in Phase2 (shocking/thunderclap)** | | | | |
| 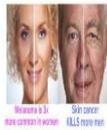 | @xyz 2 more weeks to reach goal of 100 #KnowYourSkinNI supporters! Do ur bit for #skincancer; | 11,740 | 94 | 1 |
| **Most engaging tweet in Phase2 (shocking/disgust)** | | | | |
| 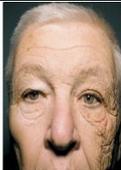 | Do you drive for a living? This is an image of a truck driver's sun damaged skin. Sunscreen & Shades! #KnowYourSkinNI  *** PROMOTED***eek | 304 | 2,655 | 3 |
| **Most retweeted message in Phase1 (Informative)** | | | | |
| 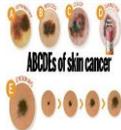 | Do you know the ABCDEs of skin cancer. A – asymmetry; B – border; C – Colour; D – Diameter... #SkinSmartNI #info | 2,258 | 100 | 17 |
| **Most retweeted message in Phase2 (Humorous)** | | | | |
| 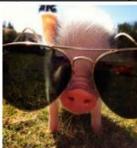 | Dust off the sunglasses, it's another lovely day in Northern Ireland. Sunscreen to hand too :) #KnowYourSkinNI | 6,367 | 196 | 13 |

**Results**

In summary, the first phase of the study generated 1,404 interactions comprising tweets, retweets and replies from 366 distinct users while the second phase generated 486 interactions from 217 distinct users. 70% of the messages were sent by users based in the UK. We inferred gender for 65% of the users using "*twitterreport*" R package. For messages on Twitter we measure message performance in terms of impressions (views) and engagements (clicks). In Table 2 we see the most retweeted messages were "*informative*" and "*humorous*" for phases 1 and 2 respectively. We also found no significant difference between promoted and non-promoted messages on both platforms.

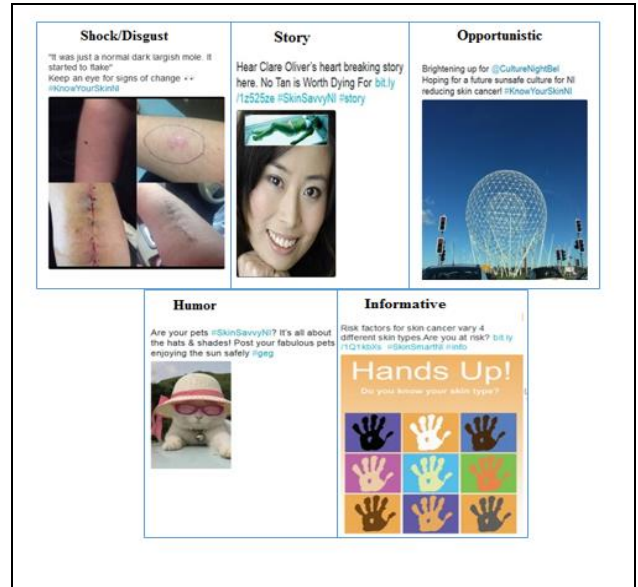

**Figure 1. Examples of message types used during the campaign**

## Future work

In our ongoing work we examine diffusion of information based on the message topic and the locations of users who propagate the information. Also, we are assessing how the various message types differ in terms of their diffusion. It would be beneficial for assessing SM enabled public health campaigns if finer granularity were obtained using a location inference algorithm (Ajao, Hong, & Weiru, 2015) which may give more location detail on campaign responses at city-level. In addition it would be interesting if future work could accurately infer more demographic characteristics of responders in platforms such as Facebook especially when response volumes were low. These features are crucial in measuring effectiveness of public health interventions.